\documentclass[prl,aps,floats,twocolumn,graphics,amsmath,amssymb,superscriptaddress,showpacs,pre]{revtex4-1}
\usepackage{graphics,graphicx}
\begin{document}

\title{The deceiving $\Delta'$: on the equilibrium dependent dynamics of nonlinear magnetic islands}

\author{F. Militello} 
\affiliation{CCFE, Culham Science Centre, Abingdon, Oxon, OX14 3DB, UK}
\author{D. Grasso} 
\affiliation{Istituto Sistemi Complessi - CNR, Torino, Italy and Politecnico di Torino, Dipartimento Energia, Torino, Italy}
\author{D. Borgogno}
\affiliation{University of Pisa, Physics Department,  Pisa, Italy}

\begin{abstract}

The linear stability parameter $\Delta'$ is commonly used as a figure of merit for the nonlinear dynamics of the tearing mode. It is shown, through state of the art numerical simulations, that factors other than $\Delta'$ can play a very important role in determining the evolution of nonlinear magnetic islands. In particular, two different equilibria are analysed and it is shown that, once perturbed, they have a qualitatively and quantitatively different response despite the fact that they are characterised by the same $\Delta'$. The different behaviour can still be associated with linear properties of the equilibrium. It is also studied how the nonlinear and saturation phase are affected by an increasing $\Delta'$ in the two equilibria. As the instability drive is increased, the systems move from a dynamics characterised by a "universal" generalised Rutherford equation to a Y-point configuration and then to a plasmoid unstable Y-point. Finally, for even larger $\Delta'$ the second harmonic overcomes the fundamental, leading to an interesting double island structure.

\end{abstract}

\maketitle

Magnetic reconnection plays a crucial role in redistributing energy in plasmas and is responsible for a variety of instabilities \cite{Biskamp2000}. In particular, magnetically confined plasmas for fusion research are subject to the tearing modes \cite{Furth1963}, which rearrange the desired axisymmetric magnetic topology by forming macroscopic magnetic islands, which degrade the confinement by increasing the cross field transport and therefore lower the machine performances. Tearing modes (TMs) and their counterpart in tokamak relevant toroidal geometry, the neoclassical tearing modes (NTMs), were recognised as a threat since the beginning of fusion research and thoroughly studied. While their linear features are relatively well understood, several aspects of their nonlinear behaviour are still elusive. The parameter $\Delta'$ determines the linear instability drive and embodies the effect of the equilibrium on the perturbation. This quantity also enters the Rutherford equation \cite{Rutherford1973} and its generalizations, which are the backbone of nonlinear tearing mode theory and modelling for weakly unstable configurations. The parameter $\Delta'$ is the \textit{de facto} figure of merit for the magnetic island evolution and its calculation in several geometrical configurations was subject of several studies \cite{Furth1973,Connor1988}.        

In this work, we show that even in the simplest magnetic reconnection models, the knowledge of $\Delta'$ does not suffice to fully determine the dynamics of the system. In particular, we will show that perturbations in different magnetic equilibria, but with the same $\Delta'$, have completely different nonlinear evolution. We have identified four instances in which this is true, corresponding to different values of the linear stability parameter (from less unstable to more unstable). 

Our work is carried out in a slab geometry and the physics is governed by the normalized reduced MHD equations \cite{Strauss1976}:
\begin{eqnarray}
\label{1} &&\frac{\partial \psi}{\partial t} +[\phi,\psi] = \eta \nabla_\perp^2 (\psi-\psi_{eq}), \\
\label{2} &&\frac{\partial \nabla_\perp^2 \phi}{\partial t} +[\phi,\nabla_\perp^2\phi] = -[\psi,\nabla_\perp^2\psi],
\end{eqnarray}
where $\psi$ is the poloidal magnetic flux, such that the magnetic field is $\textbf{B}= B_z\textbf{e}_z+\textbf{e}_z\times \nabla_\perp \psi$, the parallel current is $J=-\nabla_\perp^2\psi$, and $\phi$ is the electrostatic potential, so that $\nabla_\perp^2\phi$ is the parallel component of the $\textbf{E}\times\textbf{B}$ vorticity. The Poisson bracket $[f,g]\equiv\partial_x f\partial_y g-\partial_y f\partial_x g$ represents $\textbf{E}\times\textbf{B}$ advection of $g$ when $f=\phi$ and parallel derivation of $g$ when $f=-\psi$ ($x$ and $y$ are the "radial" and "poloidal" coordinates). The only dimensionless parameter in the problem is $\eta$, which is the normalized resistivity, equal to the inverse of the Lundquist number. In all our simulations we take $\eta=2.8\times 10^{-4}$. Note that the time is normalized with the Alfven time and the lengths with the magnetic shear length, $L_s$. 

The equations are solved numerically in a 2D domain with "radial" width $L_x=22.6$ and $L_y=2\pi/k$ where $k$ is an aspect ratio factor which can be varied to determine how unstable are the equilibria. The large value of $L_x$ allowed us to reach sizeable $\Delta'$ (i.e. big islands) without incurring in boundary effects. The resolution is 2048 points in $x$ with non equispaced grid (minimum $\Delta x \approx 4.5\times 10^{-3}$) employing compact finite differences \cite{Lele1992} and between 512 and 4096 in $y$ (depending on $L_y$), where the code is pseudo-spectral and parallel. We assume $\phi=\tilde{\psi}=0$ at $x=\pm L_x/2$ and periodicity in the $y$ direction. Two magnetic equilibria are studied \cite{Harris1958,Porcelli2002}: A) $\psi_{eq}= (3\sqrt{3}/4)/\cosh(x)^2$ and B) $\psi_{eq}= -\log[\cosh(x)]$. In both cases, we take $\phi_{eq}=0$, corresponding to no equilibrium flows.

We define $\displaystyle \Delta'_n\equiv \lim_{x \to 0}\tilde{\psi}_n^{-1}(\partial_x\tilde{\psi}_n|_{+}-\partial_x\tilde{\psi}_n|_{-})$, where $\tilde{\psi}_n$ is the order $n$ Fourier component of $\tilde{\psi}$ and the derivatives are taken to the right and to the left of the resonant surface. Note that $\Delta'=\Delta'_1$, i.e. the stability parameter of the fundamental harmonic. Using the previous definition, equilibrium A has: $\Delta'_n = 2(5 - n^2k^2)(3 + n^2k^2)/(n^2k^2\sqrt{4 + n^2k^2})$ and equilibrium B: $\Delta'_n = 2[(nk)^{-1}-nk]$. A crucial difference between the equilibria is that for the same $\Delta'$, A has $\Delta'_{n>1}$ less unstable than B. 

\begin{figure}
\includegraphics[height=6cm]{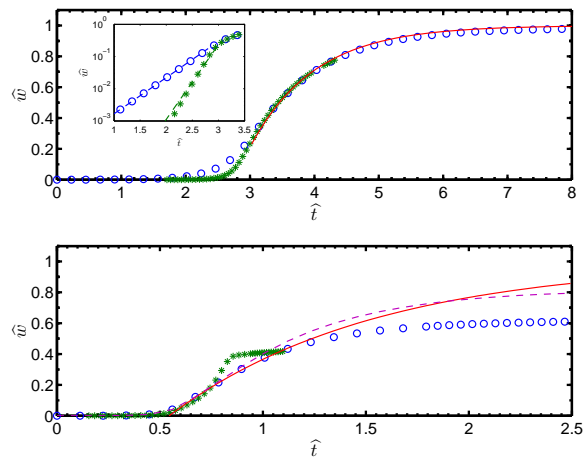} \caption{Upper panel: linear and nonlinear evolution of the magnetic island width with $\Delta'=1.23$ for equilibrium A (circles) and B (stars). The "universal" solution is shown as a solid line. In the box: linear phase in logarithmic scale compared with the theoretical predictions \cite{Militello2003}. Lower panel: same as upper panel with $\Delta'=8.1$. The dashed line is $\alpha w_a/\Delta'$.}
\label{fig1}
\end{figure}

We start with weakly unstable cases, in which $\Delta'=1.23$ for both equilibria and all the $\Delta'_{n>1}$ are negative (i.e. the higher harmonics are linearly stable). In this regime, Rutherford equation is valid but, in order to reproduce the numerical results, needs to be corrected with the saturation term rigorously derived in \cite{Militello2004,Escande2004}. Using the appropriate normalization, the equation becomes "universal", in the sense that it can be simply expressed as:
\begin{equation}
\label{3} \frac{\partial \widehat{w}}{\partial \widehat{t}}=1-\widehat{w},
\end{equation}
where $\widehat{w}=\alpha w/\Delta'$, $w=[|\tilde{\psi}_1(0)|/J_{eq}(0)]^{1/2}$ is the island width, $\widehat{t}=1.22\alpha \eta t$ and $\alpha = -0.41[J_{eq}''(0)/J_{eq}(0)]$. The solution of Eq.\ref{3} is $\widehat{w}=1+(\widehat{w}_0-1)e^{-(\widehat{t}-\widehat{t}_0)}$, where $\widehat{w}_0=\widehat{w}(\widehat{t}_0)$ is an integration constant representing the island width at the beginning of the nonlinear phase and reduces to the "standard" Rutherford equation for $\widehat{w}\ll 1$, i.e. when the island's width is much smaller than its saturation value. Differently from the "standard" solution, the "universal" solution clearly shows that the complete dynamics depends on the combination $\Delta'/\alpha$ rather than $\Delta'$ alone. To confirm this, we have performed simulations the results of which are shown in the upper panel of Fig\ref{fig1}. The linear phase, magnified in the box in the upper panel and compared with the theoretical results in \cite{Militello2003}, is followed by the nonlinear regime for $\widehat{t}>3$, which perfectly matches the "universal" solution. Note that the linear growth explicitly depends on $k$ while the nonlinear dynamics does not [only through $\Delta'(k)$]. 

\begin{figure}
\includegraphics[height=6cm]{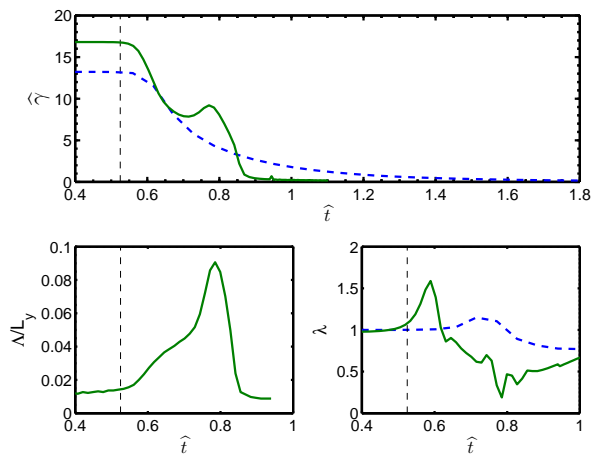} \caption{Upper panel: Comparison between the instantaneous growth rates for equilibrium A (dashed line) and B (solid line). The thin vertical dashed line marks the end of the linear phase. Left lower panel: length of the Y-point, $\Lambda$, as a fraction of the "poloidal" box size for equilibrium B. Right lower panel: opening parameter, $\lambda$, for equilibrium A (dashed line) and B (solid line).}
\label{fig2}
\end{figure}

\begin{figure}
\includegraphics[height=3cm,width=4.25cm]{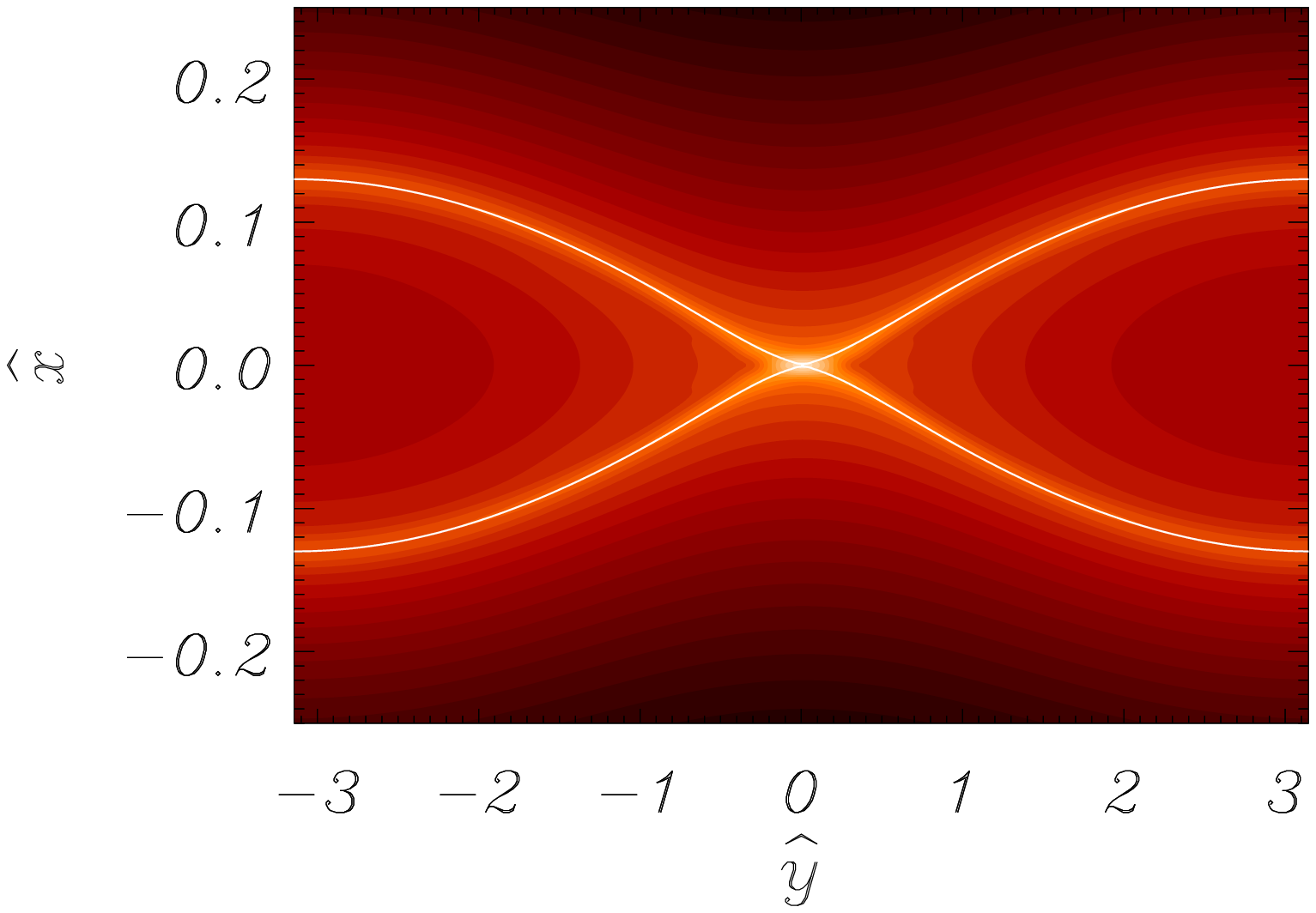}
\includegraphics[height=3cm,width=4.25cm]{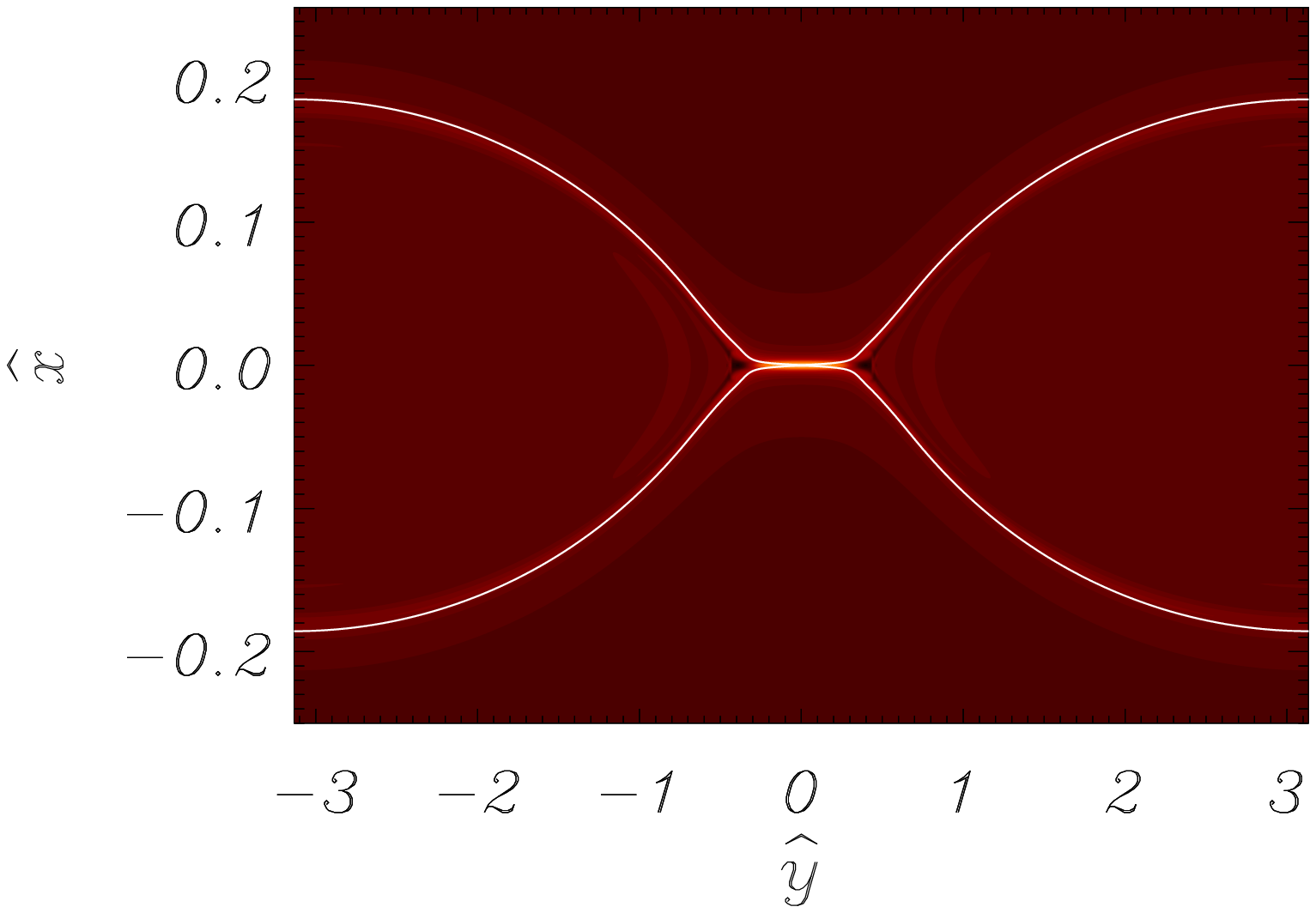}
\includegraphics[height=3cm,width=4.25cm]{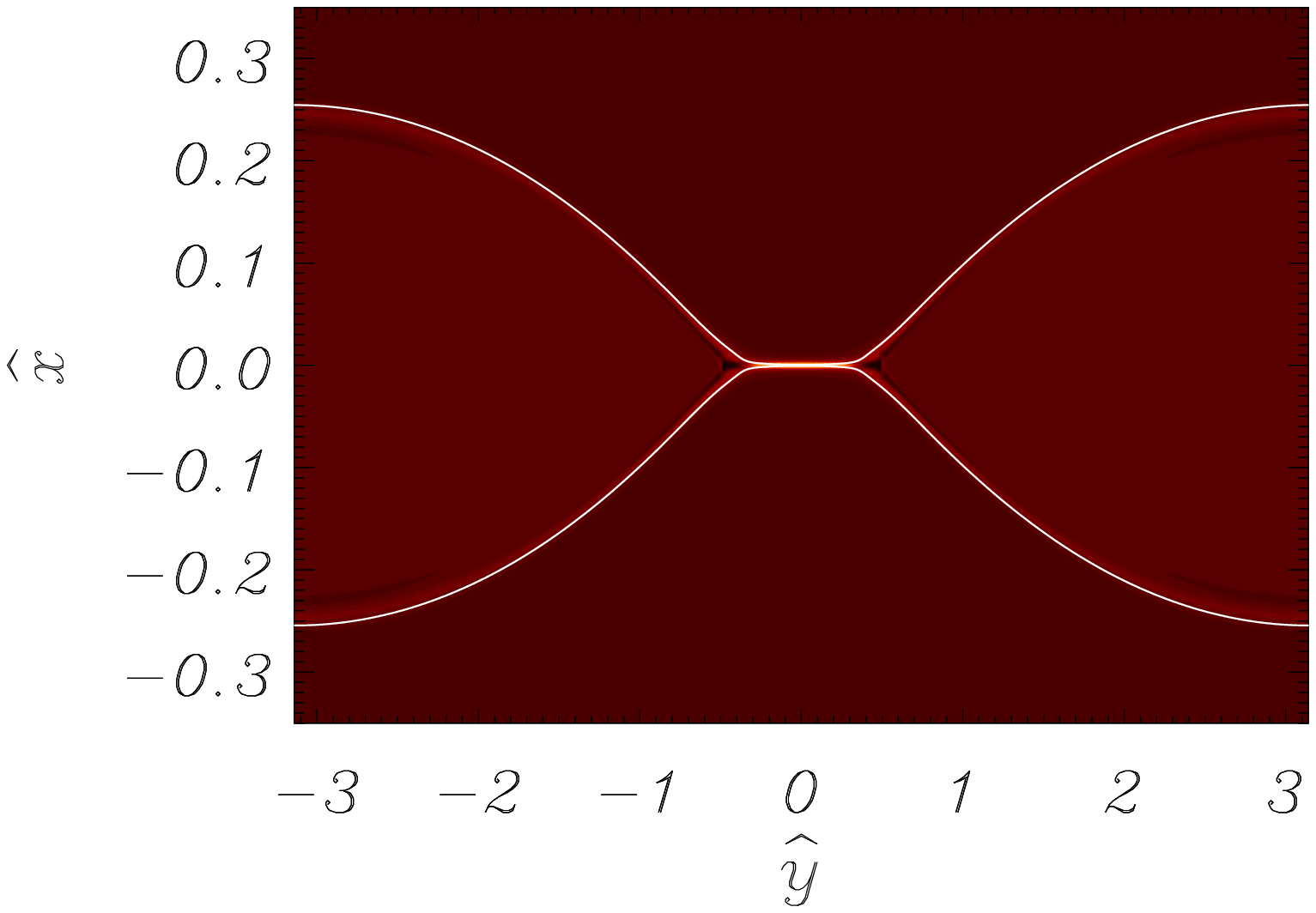}
\includegraphics[height=3cm,width=4.25cm]{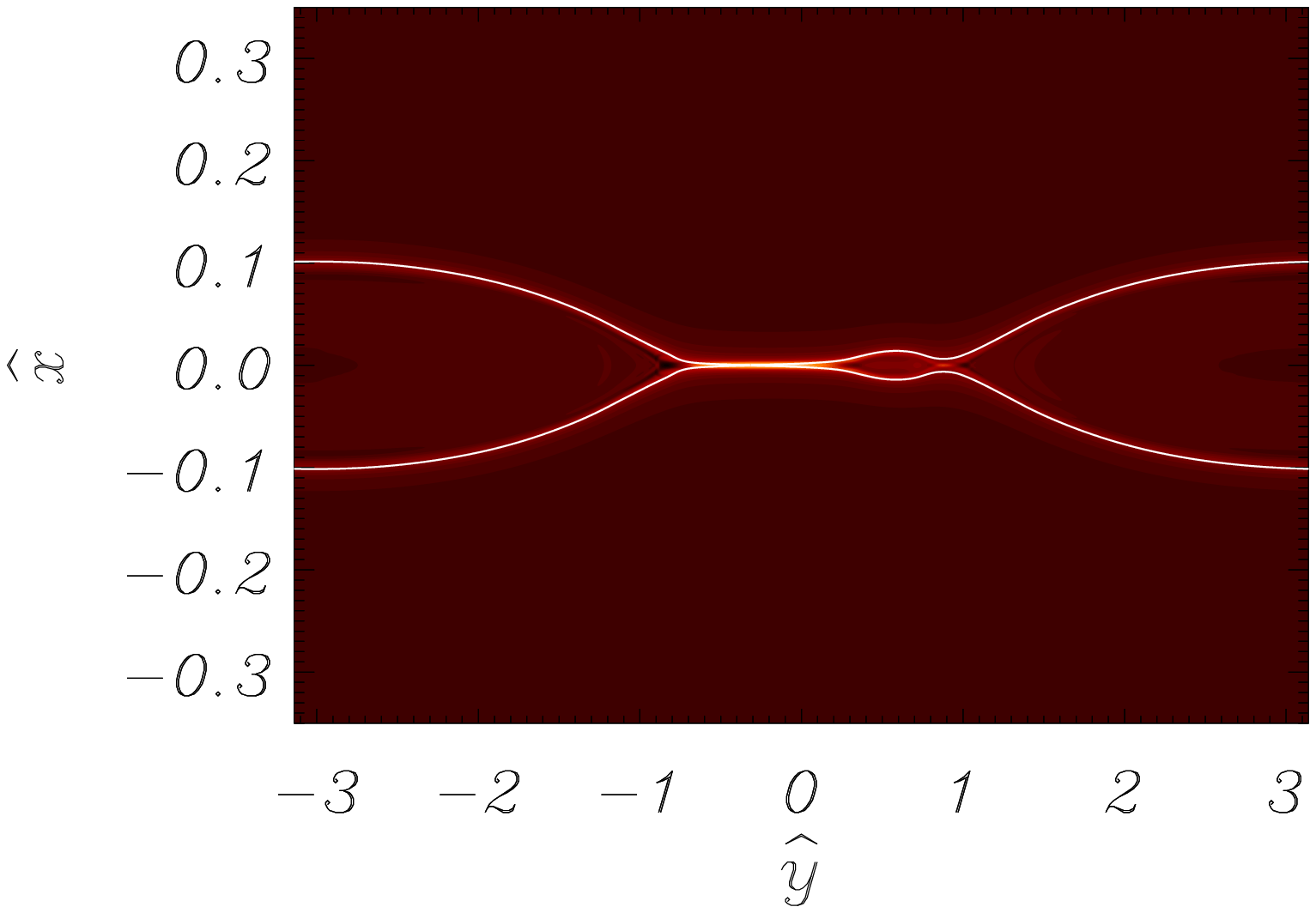}
\caption{Contour plots of the current density and separatrix shape (thin line) for $\Delta'=8.1$ at $\widehat{t}=0.78$ for equilibrium A (upper left) and B (upper right) and at $\Delta'=14.3$ for equilibrium A (lower left) and B (lower right). Note: $\widehat{x}\equiv x\alpha/\Delta'$ and $\widehat{y}=yk$.}
\label{fig3}
\end{figure}

Next, we destabilized the equilibrium by decreasing $k$ in order to have $\Delta'=8.1$ for both equilibria. In this case, the nonlinear island evolution showed  qualitative and quantitative differences for the two equilibria. In particular, equilibrium A evolves following the "universal" equation until the island reaches a macroscopic width for which $J_{eq}(w)$ is not well approximated by its local Taylor expansion $J_{eq}(0)+J_{eq}''(0)w^2/2$, one of the assumption at the base of the derivation of Eq.\ref{3} \cite{Rutherford1973,Militello2004}. The actual island width would be better represented by $w_a=2\tanh^{-1}[\sqrt{|\tilde{\psi}_1(0)|/J_{eq}(0)}]$ in this regime ($w_a=2\cosh^{-1}[e^{2|\tilde{\psi}_1(0)|/J_{eq}(0)}]$ for equilibrium B), which correctly reduces to the expression below Eq.\ref{3} for $|\tilde{\psi}_1(0)|/J_{eq}(0)\ll 1$. For the same value of $\tilde{\psi}_1(0)$, this new expression gives a larger island width than $w=[|\tilde{\psi}_1(0)|/J_{eq}(0)]^{1/2}$. This explains why Ref.\cite{Loureiro2005}, which measured $w_a$ but used the old definition of $w$ to reconstruct $|\tilde{\psi}_1(0)|$, found good agreement between the POEM saturation theory predictions and their numerical results at $\Delta'=8.1$ (see dashed line in the bottom panel of Fig.\ref{fig1}) and we do not. Equilibrium B behaves in a remarkably different way as it shows a fast nonlinear growth corresponding with the formation of a Y-point \cite{Waelbroeck1989,Waelbroeck1993,Jemella2003}. A comparison of the nonlinear phases is given in the lower panel of Fig.\ref{fig1} and the instantaneous growth rates $\widehat{\gamma} = \partial_{\widehat{t}} \log(\widehat{w})$ for the two equilibria are shown in Fig.\ref{fig2}. 
Noticeably, while reconnection in equilibrium A occurs in a X-point structure throughout the whole evolution of the island, equilibrium B forms a current ribbon, which reaches its maximum length ($\Lambda$ is 10\% of $L_y$ at $\widehat{t}\approx 0.8$) in correspondence to the peak of the instantaneous growth rate (see Figs.\ref{fig2} and \ref{fig3}) and then shrinks, yielding an X-point saturation.

The interpretation of these results is straightforward if one notices that the current ribbon can be sustained only when higher order harmonics are sufficiently large with respect to the fundamental. Expressing the magnetic flux as $\psi=\psi_{eq}(x)+\displaystyle \sum_{n=0}^\infty|\tilde{\psi}_n(x,t)|\cos(nky+\beta_n)$ and Taylor expanding it close to $x=y=0$ gives:
\begin{equation}
\label{4} \chi\approx\frac{x^2}{2}-\displaystyle \sum_{n=0}^\infty n^2\Psi_n\frac{y^2}{2}+\displaystyle \sum_{n=0}^\infty n^4\Psi_n\frac{y^4}{24}+\cdots
\end{equation}   
where $\chi\equiv[\psi-\psi_{eq}(0)]/\psi_{eq}''(0)$, $\Psi_n\equiv \sigma_n|\tilde{\psi}_n(0)|/\psi_{eq}''(0)$ and $\sigma_n\equiv\cos(\beta_n)$ can only be $\pm 1$ (see \cite{Arcis2009}). The opening angle of the X-point is therefore: $\alpha = 2\arctan(\lambda w)$, where $\lambda\equiv [\displaystyle \sum_{n=0}^\infty n^2\sigma_n|\tilde{\psi}_n(0)|/|\tilde{\psi}_1(0)|]^{1/2}$, and for vanishing $\lambda$ it goes to zero while the island retains a finite size (i.e. it forms a current ribbon). For $\lambda<0$, Eq.\ref{4} goes from hyperbolic to elliptic, i.e. the X-point is replaced by a secondary O-point. In our simulations $\lambda$ is always around unity for equilibrium A, while it becomes very small ($\sim 0.1$) when the Y-point is forming in Equilibrium B (both cases at $\Delta'=8.1$). We therefore observe that the current ribbon associated with equilibrium B has the structure of a very narrow X-point which suddenly expands at $y=\pm\Lambda/2$ (see Fig.\ref{fig3}). 

The reason why the two systems have a dissimilar evolution is that the equilibria have different stability properties for the high order harmonics. In particular, equilibrium A has $\Delta'_{n>1}$ negative, while equilibrium B has $\Delta'_2=3.38$, $\Delta'_3=1.48$ $\Delta'_4=0.299$ and $\Delta'_{n>4}<0$. In other words, equilibrium B has 4 linearly unstable harmonics. The $n>1$ linear stability parameters affect also the nonlinear theory by providing a further destabilization (if positive) on top of the nonlinear mode coupling, as already suggested in Ref.\cite{Arcis2009}. The linear drive of the secondary harmonics allows them to reach a larger size compared to the fundamental, so that the parameter $\lambda$ can become small, provided that the $\sigma_n$ have the right sign (a spectral analysis confirms these hypothesis). Note that in $\lambda$ each harmonic is weighted with respect to its mode number squared, so that even high order harmonics can play an important role, despite their small amplitude. 

The next step is to further destabilize the equilibria by taking $\Delta'=14.3$. In this case, we have $\Delta'_2=1.23$ and $\Delta'_{n>2}<0$ for equilibrium A, while 8 harmonics are unstable for Equilibrium B. This time, we expect current ribbon formation in the first configuration as well as in the second. While this is confirmed by the simulations (see Fig.\ref{fig3}), the systems remain dissimilar from a qualitative point of view. Indeed, equilibrium B now forms plasmoids \cite{Loureiro2007,Loureiro2013} which transiently appear in the region of the current ribbon. Interestingly, their generation is correlated with the condition that $\lambda<0$, contributing to suggest that also the plasmoids could find their drive in the linear structure of the equilibrium.
    
\begin{figure}
\includegraphics[height=3cm,width=8.5cm]{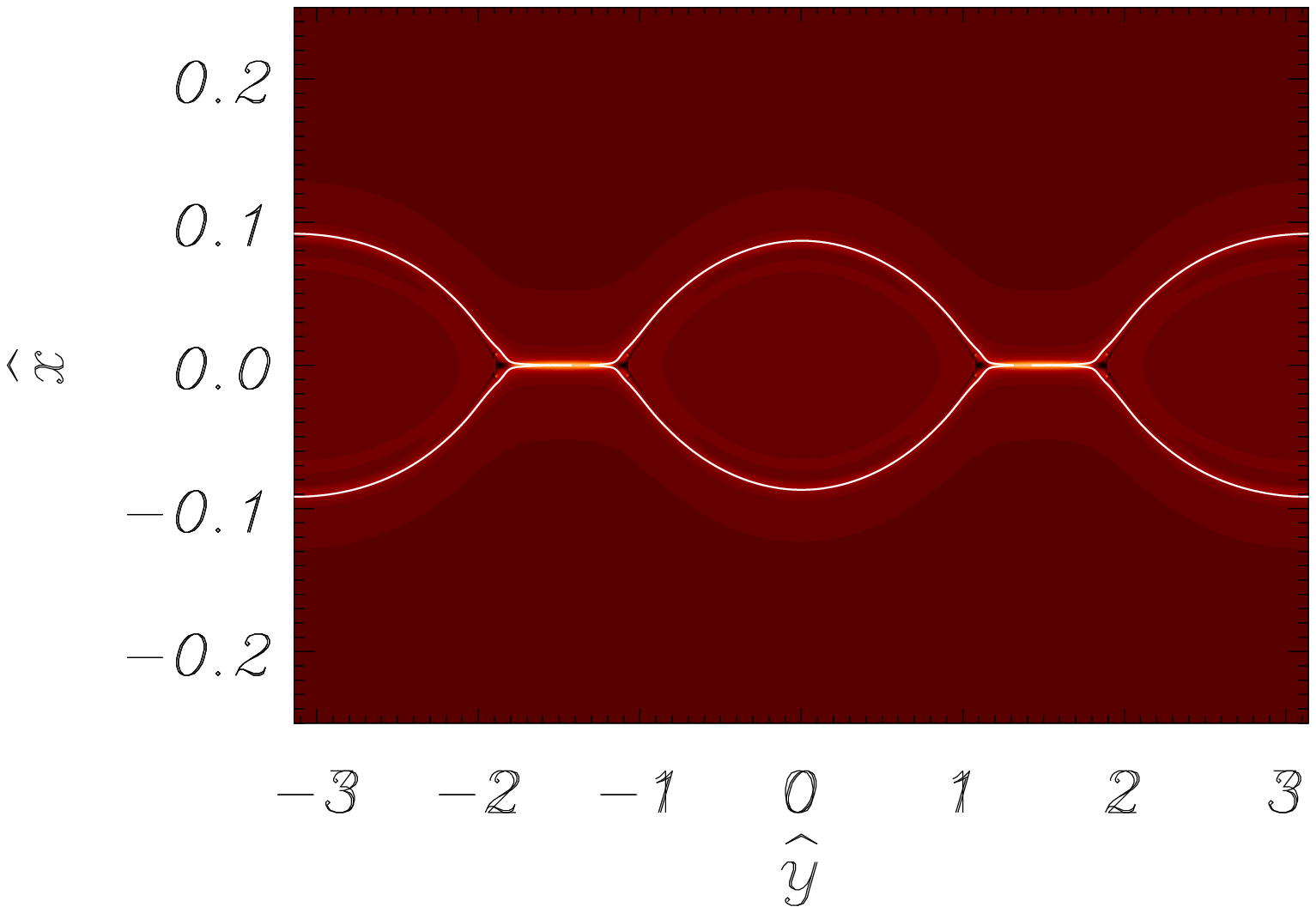}
\includegraphics[height=3.5cm,width=8cm]{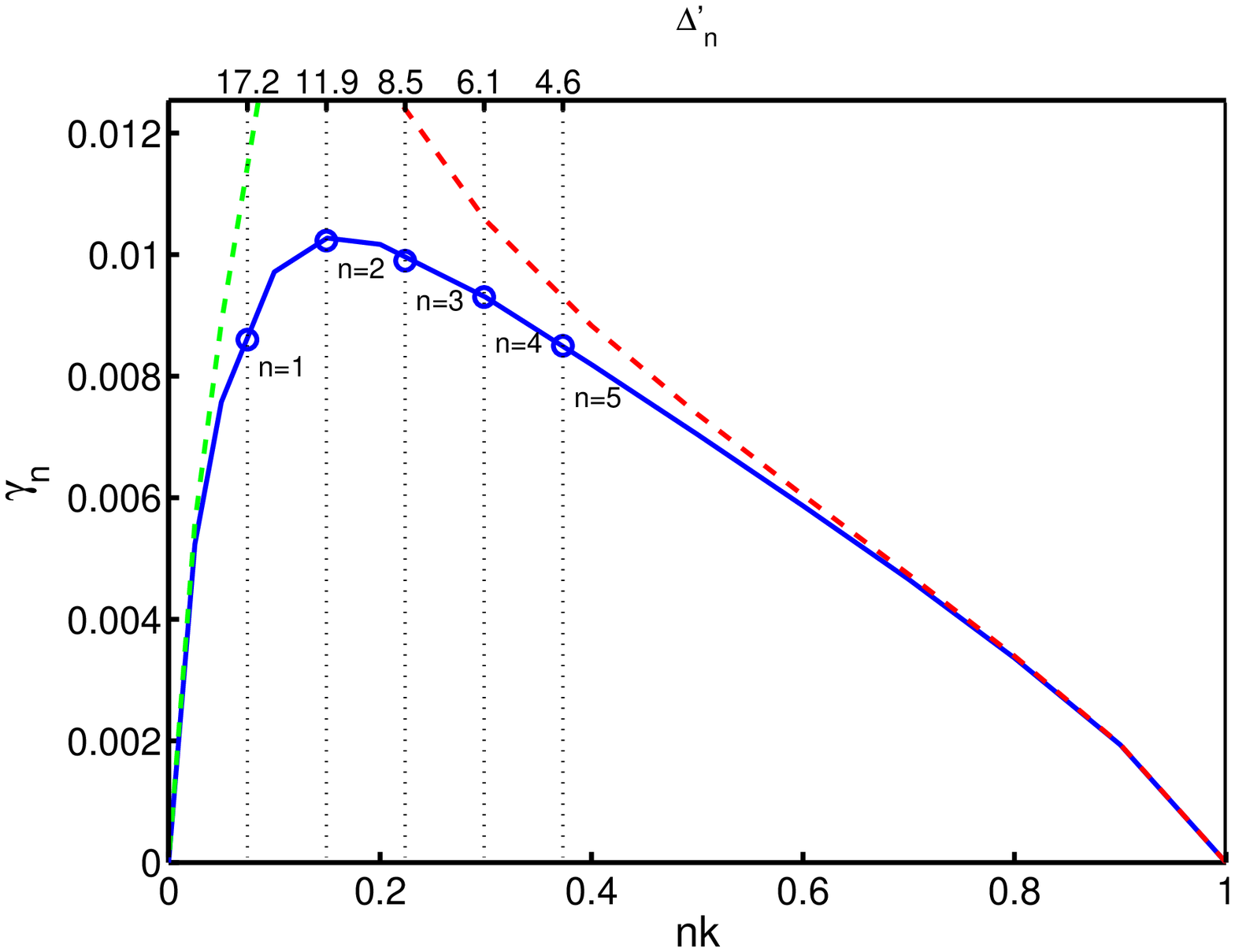} \caption{Upper panel: contour plot of the current density and separatrix shape (thin line) for equilibrium B at $\Delta'=17.25$. Lower panel, linear growth rate, $\gamma_n$ as a function of $nk$ (and $\Delta'_n$ on the upper axis)for equilibrium B. The markers show the values of the $\gamma_n$ corresponding to equilibrium B at $\Delta'=17.25$ for the first 5 modes. The dashed lines represent the asymptotic limits given in the text.}
\label{fig4}
\end{figure}

We finally run a set of simulations with $\Delta'=17.25$. Also in this case, the evolution of the perturbations in the two equilibria is completely different. In particular, equilibrium A formed a current ribbon and displayed a sudden increase of the instantaneous growth rate associated with it. Interestingly, the Y-point structure persisted for a long time towards saturation, while it quickly shrank and disappeared for less unstable cases (compare with equilibrium B at $\Delta'=8.1$, lower panel of Fig.\ref{fig2}). Equilibrium B, on the other hand, showed extremely peculiar dynamics as during the whole evolution, the second harmonic always dominated over the fundamental. This has the visible consequence of creating a double island structure, in which two islands are connected with each other through current ribbons (see Fig.\ref{fig4}). This surprising result can again be explained by the linear properties of the system. The linear growth rate is a function of $k$ both explicitly and implicitly through $\Delta'_n(k)$ and this dependence is direct for large $k$ (i.e. small $\Delta'$) and inverse for small $k$ (i.e. large $\Delta'$). Indeed, in the weakly unstable case, we have $\gamma_n\approx 0.54\eta^{3/5}\psi_{eq}''(0)^{2/5}(nk)^{2/5}\Delta_n'^{4/5}$ \cite{Furth1963,Biskamp2000}, where $\Delta_n'$ as a function of $nk$ is given above Eq.\ref{3}, and in the large $\Delta'$ regime $\gamma_n\approx \eta^{1/3}\psi_{eq}''(0)^{2/3}(nk)^{2/3}$ \cite{Ara1978,Biskamp2000}. Hence $\gamma_n$ has a maximum for $(nk)_{\gamma_{max}}\sim \eta^{1/4}\psi_{eq}''(0)^{-1/4}$ (see \cite{Biskamp2000}), so that, as we decrease $k$ to destabilize further the system, the condition $\gamma_1<\gamma_2$ can be satisfied (see Fig.\ref{fig4}). In other words, in the linear phase the second harmonic grows faster and bigger than the fundamental and this remains unaltered in the nonlinear regime, with the consequence that the macroscopic perturbation shows a double island structure. Interestingly, this gives a rule to set an upper limit to $\Delta'$ in realistic plasmas, as observing only the fundamental (e.g. with Thomson scattering) implies that $\Delta'<\Delta'(k_{\gamma_{max}})$. Note that non monotonic $\gamma_n(k)$ are found also in collisionless regime \cite{Comisso2013}. 

From the results above, we conclude that the linear stability parameter of the fundamental harmonic, $\Delta'$, is not the only linear property of the system that affects the nonlinear dynamics of the tearing modes. For weakly unstable equilibria, evaluating $\alpha$ is as important as evaluating $\Delta'$ (in asymmetric cases, also $J_{eq}'$ must be determined \cite{Hastie2005,Militello2006}).For slightly more unstable cases, the Rutherford approach must be integrated with equations that track the behaviour of the higher order harmonics (see e.g. \cite{Arcis2009}) and corrected to take into account relatively large islands. As the equilibrium gets more unstable, Rutherford's approach is entirely inappropriate as all the harmonics must be followed (for our equilibrium B, this already occurs for a moderate $\Delta'=8.1$). As the saturated island width scales with $\Delta'$, islands that cover a significant fraction of the minor radius must have a large stability parameter and hence be unsuitable to the standard modelling based on Rutherford's equation. This leads to the conclusion that, in many cases, modelling would require the solution of 2D equations, e.g. extensions of Eqs.\ref{1}-\ref{2}, rather than 0D reductions like Rutherford's equation (as supported also by \cite{Poye2013}). In addition, our work shows that the current ribbon formation is strongly dependent on the details of the equilibrium, so that it is not possible to formulate a general criterion for its occurrence based on $\Delta'w$ alone (compare with \cite{Waelbroeck1993,Loureiro2005}). From the numerical simulations and from our interpretation of the results, a necessary (but not sufficient) condition for the collapse of the X-point is $\Delta'_2>0$. Finally, we employed a very simple physical model in our study: we neglected neoclassical physics, polarisation, curvature and diamagnetic effects. However, all these terms scale in the generalized Rutherford equation as negative powers of $w$ (see \cite{Sauter1997,Waelbroeck2009} and references therein), so that they are crucial for the seed island problem, but are much less important for the large island dynamics and saturation. Similarly, the slab configuration is appropriate to capture the fundamental mechanisms governing the problem, since more complicated geometries would only affect the values of the stability parameters but they would not change their effect on the system due to the multi-scale nature of the problem \cite{Connor1988}.     

The authors acknowledge useful discussions with L. Comisso. This work was part-funded by the RCUK Energy Programme [grant number EP/I501045] and by the European Union's Horizon 2020 research and innovation programme. To obtain further information on the data and models underlying this paper please contact PublicationsManager@ccfe.ac.uk.  The views and opinions expressed herein do not necessarily reflect those of the European Commission.

\end{document}